%% file: amain.tex
\documentclass[final,conference,twocolumn]{IEEEtran}

\input{packages}
\usepackage{subfig}
\usepackage[linesnumbered,ruled,lined]{algorithm2e}

\input{acronyms2.tex}

\mathtoolsset{showonlyrefs}

\IEEEoverridecommandlockouts

\begin{document}

\input{title}
\input{notation}

\input{./1_introduction}

\input{./2_sys_model}

\input{./3_gain}
\input{./4_Asymp_analysis}

\input{./5_results}

\input{./6_conclusions}

\bibliographystyle{IEEEtran}
\bibliography{IEEEabrv,references}

\end{document}

%% file: packages.tex
\usepackage{graphicx}
\usepackage{cite}
\usepackage{amsthm}
\usepackage{amsmath,amssymb,amsfonts}
\usepackage{mathrsfs}
\usepackage{mathtools}
\usepackage{bm}
\usepackage{enumitem} 
\usepackage{cases}
\usepackage{empheq}
\usepackage{xcolor}
\usepackage{acro}
\usepackage{mathtools}
\usepackage{subcaption}
\usepackage{caption}

%% file: acronyms2.tex
\DeclareAcronym{AWGN}{short = AWGN ,long = additive white Gaussian noise}
\DeclareAcronym{ACRDA}{short = ACRDA ,long = asynchronous contention resolution diversity ALOHA}
\DeclareAcronym{CDF}{short = CDF ,long = cumulative distribution function}
\DeclareAcronym{CRA-CC}{short = CRA-CC ,long = CRA-convolutional code}
\DeclareAcronym{CRA-SH}{short = CRA-SH ,long = CRA-shannon bound}
\DeclareAcronym{CRA}{short = CRA ,long = contention resolution ALOHA}
\DeclareAcronym{CRDSA}{short = CRDSA ,long = contention resolution diversity slotted ALOHA}
\DeclareAcronym{CRDSA++}{short = CRDSA++ ,long = contention resolution diversity slotted ALOHA++}
\DeclareAcronym{CRI}{short = CRI ,long = contention resolution interval}
\DeclareAcronym{CSA}{short = CSA ,long = coded slotted ALOHA}
\DeclareAcronym{CSI}{short = CSI ,long = channel state information}
\DeclareAcronym{DAMA}{short = DAMA ,long = demand assigned multiple access}
\DeclareAcronym{DSA}{short = DSA ,long = diversity slotted ALOHA}
\DeclareAcronym{DSSS}{short = DSSS ,long = direct sequence spread spectrum}
\DeclareAcronym{E_SSA}{short = E-SSA ,long = enhanced spread spectrum ALOHA}
\DeclareAcronym{ECRA}{short = ECRA ,long = enhanced contention resolution ALOHA}
\DeclareAcronym{ECRA-SC}{short = ECRA-SC ,long = ECRA selection combining}
\DeclareAcronym{ECRA-MRC}{short = ECRA-MRC ,long = ECRA maximal-ratio combining}
\DeclareAcronym{EGC}{short = EGC ,long = equal-gain combining}
\DeclareAcronym{FEC}{short = FEC ,long = forward error correction}
\DeclareAcronym{GEO}{short = GEO ,long = geostationary orbit}
\DeclareAcronym{HAP}{short = HAP ,long = high-altitude platform}
\DeclareAcronym{IC}{short = IC ,long = interference cancellation}
\DeclareAcronym{IoT}{short = IoT ,long = Internet of things}
\DeclareAcronym{IRA}{short = IRA ,long = irregular repetition ALOHA}
\DeclareAcronym{IRCRA}{short = IRCRA ,long = irregular repetition contention resolution ALOHA}
\DeclareAcronym{IRSA}{short = IRSA ,long = irregular repetition slotted ALOHA}
\DeclareAcronym{LDPC}{short = LDPC ,long = low-density parity-check}
\DeclareAcronym{LEO}{short = LEO ,long = low-Earth orbit}
\DeclareAcronym{M2M}{short = M2M ,long = machine-to-machine}
\DeclareAcronym{MAC}{short = MAC ,long = medium access control}
\DeclareAcronym{MF}{short = MF ,long = matched filter}
\DeclareAcronym{MF-TDMA}{short = MF-TDMA ,long = multi-frequency time division multiple access}
\DeclareAcronym{MRC}{short = MRC ,long = maximal-ratio combining}

\DeclareAcronym{NOMA}{short = NOMA ,long = non-orthogonal multiple access}

\DeclareAcronym{pmf}{short = pmf ,long = probability mass function}

\DeclareAcronym{PDF}{short = PDF ,long = probability density function}
\DeclareAcronym{PER}{short = PER ,long = packet error rate}
\DeclareAcronym{PLR}{short = PLR ,long = packet loss rate}
\DeclareAcronym{QPSK}{short = QPSK ,long = quadrature phase-shift keying}
\DeclareAcronym{RA}{short = RA ,long = random access}
\DeclareAcronym{RCB}{short = RCB ,long = random coding bound}
\DeclareAcronym{RTT}{short = RTT ,long = round trip time}
\DeclareAcronym{SA}{short = SA , long = slotted ALOHA}
\DeclareAcronym{SB}{short = SB ,long = Shannon bound}
\DeclareAcronym{SC}{short = SC ,long = selection combining}
\DeclareAcronym{SIC}{short = SIC ,long = successive interference cancellation}
\DeclareAcronym{SNIR}{short = SNIR ,long = signal-to-noise and interference ratio}
\DeclareAcronym{SINR}{short = SINR ,long = signal-to-interference and noise ratio}
\DeclareAcronym{SNR}{short = SNR ,long = signal-to-noise ratio}
\DeclareAcronym{TDMA}{short = TDMA ,long = time division multiple access}
\DeclareAcronym{UCP}{short = $\Code$-UCP ,long = $\Code$-unresolvable collision pattern}
\DeclareAcronym{VF}{short = VF ,long = virtual frame}

\DeclareAcronym{rv}{short = r.v., long = random variable}

\DeclareAcronym{DVB}{short =DVB, long = digital video broadcasting}

%% file: title.tex
  
 \title{Multi-Satellite NOMA-Irregular Repetition \\ Slotted ALOHA for IoT Networks} 
\author{
\IEEEauthorblockN{Estefan\'ia Recayte\IEEEauthorrefmark{1},  Carla Amatetti\IEEEauthorrefmark{2}}     \\
\IEEEauthorblockA{\IEEEauthorrefmark{1} Institute of Communications and Navigation, German Aerospace Center (DLR),  We\ss ling, Germany } 
\IEEEauthorblockA{\IEEEauthorrefmark{2} Dept. of Electrical, Electronic, and Information Engineering,  University of Bologna, Bologna, Italy}
 
\thanks{\small E. Recayte  acknowledges the financial support by the Federal Ministry of Education and Research of Germany in the programme of “Souverän. Digital. Vernetzt.” Joint project 6G-RIC, project identification number: 16KISK022. 
This work has been funded by the 6G-NTN project, which received funding from the SNS JU under the European Union’s Horizon Europe research and innovation program under Grant Agreement No 101096479. The views expressed are those of the authors and do not necessarily represent the project. The Commission is not liable for any use that may be made of any of the information contained therein. 
}
 }

 \maketitle
 


\thispagestyle{empty} \pagestyle{empty}

\begin{abstract}
As the transition from 5G to 6G unfolds, a substantial increase in Internet of Things (IoT) devices is expected, enabling seamless and pervasive connectivity across various applications. Accommodating this surge and meeting the high capacity demands will necessitate the integration of Non-Terrestrial Networks (NTNs). However, the extensive coverage area of satellites, relative to terrestrial receivers, will lead to a high density of users attempting to access the channel at the same time, increasing the collision probability. In turn, the deployment of mega constellations make it possible for ground users to be in visibility of more than one satellite at the same time, enabling receiver diversity.  Therefore, in this paper, we evaluate the impact of multi-receivers in scenarios where IoT nodes share the channel following a \ac{NOMA}-\ac{IRSA} protocol. Considering the impairments of satellite channels, we derive a lower bound of system performance, serving as a fast tool for initial evaluation of network behavior. Additionally, we identify the trade-offs inherent to the network design parameters, with a focus on packet loss rate and energy efficiency. Notably, in the visibility of only one extra satellite as receiver yields significant gains in overall system performance.
\end{abstract}




%% file: notation.tex
\newcommand{\users}{\mathsf{m}}
\newcommand{\slots}{\mathsf{n}}
\newcommand{\load}{\mathsf{G}}
\newcommand{\userDegree}{r}
\newcommand{\avgUser}{\bar{\mathsf{r}}}
\newcommand{\slotDegree}{l}
\newcommand{\area}[1]{a_{#1}}
\newcommand{\SNIR}{\mathsf{\Gamma}}

\newcommand{\puserToslot}{\mathsf{p}}
\newcommand{\qslotTouser}{\mathsf{q}}

\newcommand{\estef}{\textcolor{blue}}
\newcommand{\change}{\textcolor{orange}}

\newcommand{\pmf}{\Lambda}
\newcommand{\prob}{p}
\newcommand{\energy}{\mathsf{E}}

\newcommand{\setS}{s}

\newcommand{\p}{\mathsf{p}}
\newcommand{\q}{\mathsf{q}}
\newcommand{\lo}{\mathsf{G}}
\newcommand{\rate}{\mathsf{R}}
\newcommand{\dt}{\gamma}
\newcommand{\power}{\mathsf{P}}
\newcommand{\energytot}{\mathtt{E}}

\newcommand{\ptime}{\mathsf{T}}
\newcommand{\kreceivers}{\mathsf{k}}
\newcommand{\ec}{\mathsf{E}}
\newcommand{\E}{\mathsf{E}}
\newcommand{\plr}{\mathtt{P}_{\epsilon}}
\newcommand{\plrtot}{\mathtt{PLR}}
\newcommand{\tr}{\mathtt{T}}
\newcommand{\se}{\mathtt{S}}
\newcommand{\per}{\mathsf{P_{L}}}
\newcommand{\pdec}{\mathsf{P_{dec}}}

\newcommand{\ee}{\eta}

\newcommand{\nsl}{\mathsf{m}}

\newcommand{\numBit}{k}
\newcommand{\nsy}{\mathsf{L}}
\newcommand{\syd}{\mathsf{T_s}}

\newcommand{\noise}{\mathsf{N}}
\newcommand{\noiseVec}{\bm{\noise}}
\newcommand{\noiseVar}{\sigma_{\noise}^2}

\newcommand{\Pw}{\mathsf{p}}
\newcommand{\packetT}{\mathsf{t}}
\newcommand{\Int}{\mathsf{I}}
\newcommand{\Ns}{\mathsf{N}}
\newcommand{\fpw}{\alpha}
\newcommand{\nil}{\tau}

\newcommand{\sinr}{\gamma}

\newcommand{\mutInf}{\mathsf{I}}
\newcommand{\avMutInf}{\bar{\mutInf}}
\newcommand{\rvDec}{\mathcal{D}}

\newcommand{\us}{\mathsf{u}}
\newcommand{\rep}{\mathsf{r}}
\newcommand{\nin}{\mathsf{t}}

\newcommand{\nurv}{U}
\newcommand{\nus}{u}

%% file: 1_Introduction.tex
\section{Introduction}\label{sec:introd}
Non-Terrestrial Networks (NTN) have steadily gained momentum as a crucial component of global connectivity, complementing conventional terrestrial infrastructure by providing coverage to remote and otherwise inaccessible areas \cite{NTN, NTNbook}. With the inclusion in the third-generation partnership project (3GPP) standards starting
from Release 17, NTNs have become integral to the future 6G-Internet of Things (IoT) ecosystem.
In parallel to this, the growing commercial interest in
massive machine-type communications (mMTC) and IoT applications introduces new challenges, pushing the connectivity requirements of next-generation   networks to new limits \cite{IoTsurvey}.  A key characteristic of IoT applications is the deployment of a large population of low-power, low-complexity devices, which are active sporadically and unpredictably, transmitting small packets over a shared channel.

In this context, modern random access schemes \cite{Berioli16}, have emerged as techniques designed to meet the demands of massive connectivity without requiring centralized coordination among users. Specifically, the \acl{IRSA} (IRSA) \cite{Liva:IRSA} protocol has proven to be an effective solution for IoT applications, supporting high throughput levels and providing tangible gains. This protocol relies on packet repetition over the \ac{MAC} frame and an iterative decoding algorithm at the receiver to solve collisions, i.e., \ac{SIC}. 
Specifically, when a packet is correctly received, the interference caused by its copies can be eliminated, allowing previously collided packets to be retrieved.
Today, IRSA and its special case, \ac{CRDSA}, are used by the return channel of the \ac{DVB}-RCS2 standard \cite{DVB-IRSA}.
 
In ongoing efforts to develop more efficient access schemes for IoT, recent literature has proposed to apply \acl{NOMA} (NOMA) in the power dimension to random access. The pioneering work \cite{choi_noma-based_2017} explores a scheme in which IoT devices select from a set of predetermined power levels to transmit their packets over an ALOHA channel. By exploiting power differences, this approach enables the receiver to capture stronger packets,  increasing the amount of information received correctly. The NOMA approach has been further investigated in protocols that support power diversity, such as \ac{CRDSA} in \cite{Rama:CRDSA, Enhancing_CRDSA}, and in \ac{IRSA}\cite{huang_iterative_2021} and in \cite{Noma:shao} for satellite communications. 

Another line of research has highlighted the promising potential of using multiple receivers in IoT to exploit spatial diversity. 
Satellite channels can experience losses due to varying channel conditions between the IoT device and the satellite, compromising the system performance. To mitigate this issue, leveraging large constellations and incorporating multiple receivers can significantly enhance the overall robustness of the system. 
The impact of multiple receivers on the throughput of a slotted {ALOHA} system was investigated, for example, in \cite{munari_multiple-relay_2021} and \cite{FORMAGGIO}. Additionally, the authors in \cite{zhao_multisatellite_2019} examine how multi-satellite cooperation enhances packet detection in modern random access protocols.

Unlike previous studies in the literature, this work evaluates the impact of multiple receivers within an IoT-NTN network operating under the NOMA-IRSA scheme. We extend the findings of \cite{Noma:shao} and \cite{feng_irregular_2022} considering channel impairments along with integration of multiple receivers.  
We evaluate our network in terms of packet loss rate and energy efficiency of the IoT devices. We formulate an analysis to establish a lower bound of the decoding failure probability for asymptotically long MAC frames, providing a rapid estimation of the system behavior. Moreover, we report the outcome of detailed simulations that account for the effect of the finite frame duration.
Our study offers valuable insights into understanding the fundamental trade-offs associated with parameter design, supporting ongoing efforts to develop more sustainable and efficient IoT networks \cite{Recayte:twos}.

The rest of the paper is organized as follows. Section~\ref{sec:sysmodel} introduces the system model, while Section~\ref{sec:plr} describes the evaluation of the performance metrics. The asymptotic analysis is presented in Section~\ref{sec:Asym}. Numerical results are provided in Section~\ref{sec:results}, and finally Section~\ref{sec:Conclusions} summarizes the conclusions.

%% file: 2_sys_model.tex
\begin{center}
  \begin{figure}[t]
  \includegraphics[width=.8\columnwidth]{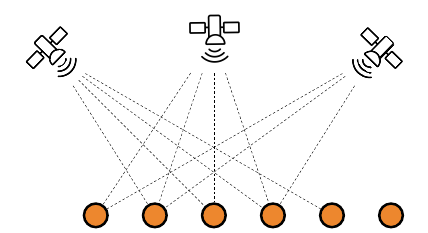}
     \centering 
     \caption{Example of $\users =6 $ IoT nodes connected to $\kreceivers= 3$ satellites. The OOF channel model is represented such that a missed link between a user and a satellite indicates an erased transmission. }
     \label{fig:system_model}
  \end{figure}   
\end{center}
\vspace{-2em}
\section{System Model}\label{sec:sysmodel} 
We consider the uplink scenario sketched in Fig.~\ref{fig:system_model}  where a population $\mathcal{U}= \{\us_{1}, ..., \us_{\users}\}$ of $\users$ uncoordinated IoT users (\emph{nodes}) transmits packets to satellites over a shared medium, representing an NTN network.


\subsection{Medium access contention}
We assume that users follow the \ac{IRSA} protocol \cite{Liva:IRSA}, where time is divided into slots of fixed duration, corresponding to the packet time duration $\ptime$. The \ac{MAC} frame consists of $\slots$ consecutive slots, with both users and the receiver sharing common knowledge of the frame's start and end times, as well as the timing of each slot.\footnote{This is typically implemented with a beacon message.} The channel load, denoted as $\load$, is defined as 
\begin{equation}\label{eq:load}
  \load = \frac{\users}{\slots} \quad [\text{packets/slot}].
\end{equation}

Each IoT node transmits $\ell$ replicas of its data packet,  distributed uniformly at random across the $\slots$ available slots within the frame. The number of replicas $\ell$ is independently drawn by each user according to a predefined  probability mass function. Following the traditional polynomial formulation, this can be expressed as \cite{Liva:IRSA}
\begin{equation}\label{eq:IRSAdist}
  \Lambda(x) = \sum_{\ell=2}^{\ell_{\max}} \Lambda_{\ell} \, x^\ell
 \text{ with } \sum_{\ell=2}^{\ell_{\max}} \Lambda_{\ell}  = 1
\end{equation}
where $\Lambda_\ell$ represents the probability that a node transmits $\ell$ replicas, while $\ell_{\max}$ denotes the maximum number of replicas a node can send within a frame. For instance, the \ac{IRSA} distribution ${\Lambda(x) =  0.25 x^2 + 0.75 x^3}$ indicates that a user  sends two replicas with  probability  0.25 and three with  probability 0.75. We also define the average number of transmitted replicas per user as  $\bar{\ell} =   \sum_{\ell=2}^{\ell_{\max}}\ell  \Lambda_{\ell} $. 

Each replica includes the indexes of the slot where its twins are located. This information can be obtained, for example, by inserting pointers into the packet header.\footnote{A  efficient method is to use the packet payload as the seed for the random generator employed by the sender to place its replicas. In this way, once one replica is decoded, the receiver can determine the positions of all the others.}

\subsection{Replica power diversity}\label{sec:sys_noma}
We consider a \ac{NOMA} scheme in which each replica might be transmitted at a different power level. Once the number of replicas has been determined, each user selects a transmission power level for each of them. The IoT devices operate under a peak power constraint,\footnote{This assumption is based on practical factors,  such as the performance limitations of the power amplifiers and regulatory requirements governing spectrum usage.} denoted by $\power$.  To manage power allocation efficiently, we assume two distinct power levels, \emph{weak} and \emph{strong} as in \cite{RecayteIRSA}. These power levels are fractions of the peak power, specifically 
\begin{equation}\label{eq:p1p2} p_2 = (1 - \alpha) \power   \text{ and } p_1 = \alpha \power \end{equation} 
where $ 0< \alpha < 0.5$ and   $p_2, p_1$ represent the \emph{strong} and \emph{weak} power levels, respectively. The parameter $\alpha$ defines the proportion of peak power employed. A replica can be assigned to either strong or weak power with equal probability.

Following the approach introduced in \cite{choi_noma-based_2017}, the power levels need to satisfy the following condition
\begin{equation}\label{eq:p_k}
p_k = \gamma (\gamma + 1)^{2-k}  \quad k \in [1,2 ]
\end{equation}where $\gamma$ represents the target \ac{SINR}  necessary for the receiver to successfully decode a replica. This formulation ensures that  a replica transmitted with strong power can be decoded even in the presence of interference from a weak power transmission within a slot. Additionally, a replica transmitted at weak power can be successfully decoded in interference-free slot.

By manipulating the relationship in \eqref{eq:p_k} with  the definition of $p_1$ and $p_2$, we derive the value  
\begin{equation}\label{eq:alpha}
  \alpha = (\sqrt{\power +1} -1)/{\power}.  
\end{equation}
This expression establishes a relationship between the peak power constraint and the power levels used in the transmission, ensuring that both power levels are optimized for the network performance.

\subsection{Channel model}
To capture the significant channel impairments resulting from the short-term unavailability of an NTN link, we assume that packets transmitted by nodes are subject to erasures. Specifically, we consider the on-off fading (OOF)  channel framework proposed in \cite{Perron:oof}. 

In this model, a packet is completely erased with probability $\epsilon$, meaning it does not contribute any interference at the receiver and is effectively lost. For example, this can occur due to severe rain conditions or physical obstructions.
Alternatively, the packet arrives successfully at the receiver with probability $(1- \epsilon).$
Erasures are independent and identically distributed among users \cite{munari_multiple-relay_2021}, while it is assumed that replicas from the same user within the frame experience the same realization of the on-off channel.

Additionally, when multiple satellites are considered as receivers, each user experiences independent channel realization with each satellite. This means that replicas from a given user may be completely lost at one receiver while being successfully received by another. We illustrate this scenario in Fig.~\ref{fig:system_model}, where, for example, the first user on the left is connected to two out of the three satellites.

\subsection{Successive interference cancellation decoding}
At the receiver side, the satellite performs \ac{SIC} over a MAC frame \cite{Liva:IRSA}. This algorithm processes slots and the frame iteratively. Whenever the \ac{SINR} of a replica, denoted by $\Gamma$, exceeds the threshold $\gamma$, i.e.  $\Gamma \geq \gamma$,  the replica is decoded.  

Recalling \eqref{alg:DE}, the inequality  $\Gamma \geq \gamma$ holds whenever a replica is received without collisions within a slot.
In the NOMA scheme, the capture effect can be exploited at the receiver, meaning that collisions are not always destructive. Specifically, if a slot contains two replicas - one with higher power 
and the other with lower power - 
the receiver can first decode the stronger signal and remove it from the slot. This also allows the receiver to decode the weaker signal later.

When the receiver decodes a replica, then it has information about where its replica twins are located, and their contribution from the incoming signal is also canceled. Due to cancelations, replicas from other users may become decodable. This iterative procedure is repeated until no additional replicas with $\Gamma \geq \gamma$ can be identified.

\subsection{Multiple receiver}

We aim to evaluate the performance of the proposed scenario in the presence of multiple receivers. To accomplish this, we consider 
$\kreceivers \geq 1 $ satellites.

After $\slots$ slots, each satellite may observe a different version of the MAC frame due to independent erasures. At the end of a frame, with $\users$ users having transmitted, each satellite may experience a distinct interference pattern across slots, as different users may have been erased.  However, a correlation exists between the MAC frames, as replicas sent by a user without erasure appear in the same slots across all satellites.

Each receiver $j = {1, ..., \kreceivers}$ then independently runs the SIC algorithm, resulting in the decoding of subsets $\mathcal{D}_j \subseteq  \mathcal{U}$ of users.
A total of $d$ aggregated users are decoded by the entire system over a MAC frame, where $d$
\begin{equation}
   d = \left|\bigcup_{j=1}^{\kreceivers} \mathcal{D}_j\right|. 
\end{equation}

%% file: 3_gain.tex
\section{Multi-receiver  Performance Metrics}\label{sec:plr}
We aim to characterize the impact of assuming multiple receivers in the proposed NOMA-IRSA scenario. To achieve this, we analyze the system using two key performance metrics: packet loss rate $\plrtot$ and energy efficiency $\eta$.

\subsection{Packet loss rate}
Let us denote by $\plrtot(\load)$ the probability that a user is not decoded by the system when channel load is $\load$. 
This probability represents the likelihood that an arbitrary user, 
$\us$, is not included in any subset of users decoded by any satellite. Therefore, we can write
\begin{align} \label{eq:plrtot1}
  \!\!\!\!  \plrtot(\load)  \!\!= \Pr \!\left[ \us \not\in \bigcup_{j=1}^{\kreceivers} \mathcal{D}_j  \right] \! \geq  \!\prod_{j=1}^{\kreceivers} \Pr \left[ \us \not\in  \mathcal{D}_j  \right]   =  \plrtot_{\mathsf{B}}(\load)
\end{align}
where the inequality results from the correlation between the MAC frames observed by the satellites. Thus, $\plrtot_{\mathsf{B}}(\load)$ represents a lower bound.

The term  $\Pr \left[ \us \not\in  \mathcal{D}_j  \right]$  evaluates the probability that a user is not decoded by the 
$j$-th satellite. This occurs if the user's replicas were either not received (erased) with probability $\epsilon$, or if they were received with probability $1- \epsilon$ but the SIC algorithm failed to decode the user. Denoting the probability of the latter event by $\plr(\load)$, we can express  
\begin{equation}\label{eq:pD}
    \Pr \left[ \us \not\in  \mathcal{D}_j  \right] =  \epsilon +  (1-\epsilon) \plr(\load)  .
\end{equation}
The derivation of $\plr(\load)$ is challenging due to the complexity in tracking the behavior of the SIC decoding algorithm over finite frame length \cite{Liva:IRSA}. However, the evaluation can be performed through Algorithm 1, which outputs the value of $\plr(\load)$ considering 
asymptotically long frames. Details of its derivations are provided in the next Section.

Finally, by inserting \eqref{eq:pD} into \eqref{eq:plrtot1}, we find that
\begin{equation}\label{eq:plr}
     \plrtot_{\mathsf{B}}(\load)   = \left[   (1-\epsilon) \plr(\load)  + \epsilon \right]^\kreceivers. 
\end{equation}
Due to the binomial theorem, this can also be written as ${ \plrtot_{\mathsf{B}}(\load)   = \sum_{i=0}^{\kreceivers} \binom{\kreceivers}{i} (1-\epsilon)^{\kreceivers-i}   \plr(\load)^{\kreceivers-i}  \epsilon^i.}$

\subsection{Energy efficiency}
The energy efficiency of the system reflects the balance between energy expenditure and successful data transmission. By understanding this relationship, we can better assess the network's operational viability and identify potential optimizations for prolonging the devices lifespan.

Let $\mathsf{\bar{E}}$ denote the average energy consumed by an IoT node to transmit its replicas within a MAC frame. Given that the probability of transmitting a replica at either strong or weak power is equiprobable, we have that 
\begin{align}
\mathsf{\bar{E}} & = \bar{\ell} \Big[ \frac{1}{2} p_1 + \frac{1}{2} p_2 \Big]  \mathsf{T} \\
& = \frac{ \bar{\ell}}{2}\Big[  (1-\alpha) \, \power + \alpha \, \power \Big]  \mathsf{T}    = \frac{ \bar{\ell} \, \power \ptime}{2}.
\end{align}
Let the average number of successfully decoded users in the system per time slot, or \emph{throughput}, be denoted by ${\load \big[1 - \plrtot(\load) \big]}$. The energy efficiency of the system is then defined by the following ratio
\begin{equation}\label{eq:energy}
    \eta = \frac{ 2 \, \load \big[1 - \plrtot(\load) \big]  }{\bar{\ell} \, \power  \, \ptime }.
\end{equation}
The value of $\eta$ represents the rate of successful data transmission per energy expended. 

\SetKwInput{KwInput}{Input}                
\SetKwInput{KwOutput}{Output}
\begin{algorithm}[h]\vspace{.2em}
 \KwInput{$\load, \epsilon, \Lambda_{\ell}, \bar{\ell}, \mathsf{I_{\max}}$}\vspace{.2em}
 \KwOutput{$\plr(\load)$}\vspace{.3em}
$\q_{0} = 1 $ \\
$\p_{0} = 1 $ \\
$  i = 1 $ \\
\While{$i \leq \mathsf{I_{\max}} $}{ \vspace{.2em}
 $   \q_{i} = \sum_{\ell=1}^{\ell_{\text{max}}} \,   \Lambda_{\ell} \,  \p_{i-1}^{\ell-1}.  $\\ \vspace{.2em}
$ \p_{i} =  1 - e^{-\bar{\ell} \, \load \, (1 - \epsilon) \q_{i} } \big[1 + \frac{\bar{\ell}}{2}  \load (1 - \epsilon) \q_{i}  \big]$ \\ \vspace{.2em}
$ i = i +1 $ \\ \vspace{.2em}
}\vspace{.2em}
$\plr(\load)= \sum_{\ell=1}^{\ell_{\text{max}}} \,   \Lambda_{\ell} \,  \p_ \mathsf{{I_{\max}}}^{\ell} $ 
\caption{Density evolution to derive $\plr(\load)$ }\label{alg:DE}
\end{algorithm}

%% file: 4_Asymp_analysis.tex
\section{Asymptotic Analysis}\label{sec:Asym}
\begin{figure*}[t!]
    \centering
    \begin{subfigure}[b]{0.32\textwidth}
        \centering
        \includegraphics[width=\textwidth]{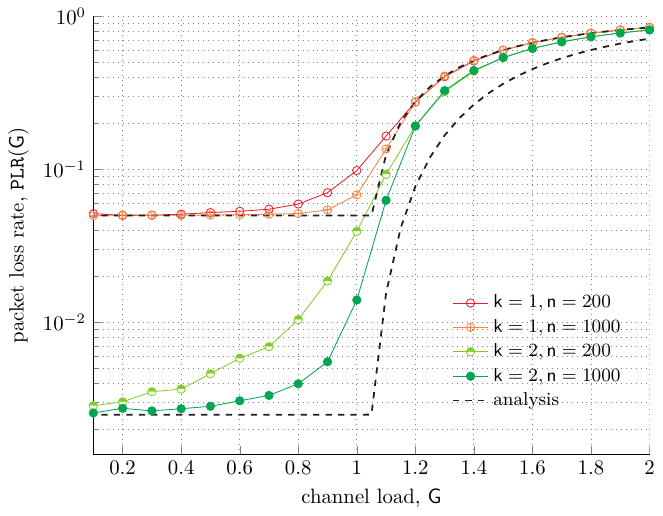}
        \caption{Degree distribution $\Lambda_1(x)$, channel erasure probability $\epsilon=0.05$ for $\kreceivers$ satellites and $\slots$ slots}
        \label{fig:DE}
    \end{subfigure}
    \hfill
    \begin{subfigure}[b]{0.32\textwidth}
        \centering
        \includegraphics[width=\textwidth]{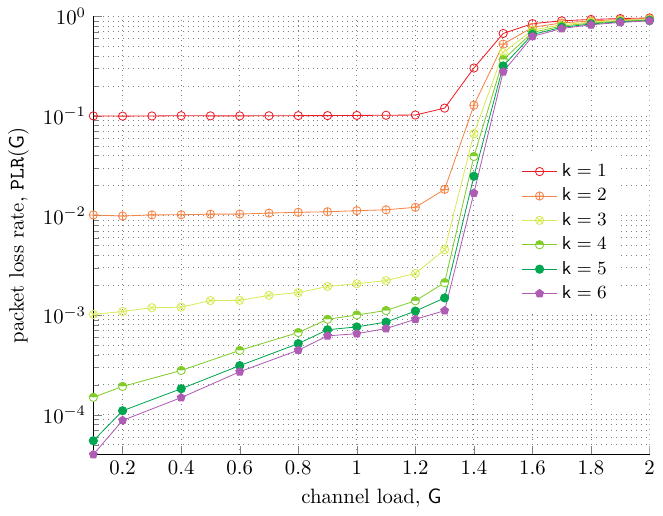}
        \caption{Degree distribution $\Lambda_2(x)$, channel erasure probability $\epsilon=0.10$, $\slots=200$ slots for $\kreceivers$ satellites}
        \label{fig:PER_k}
    \end{subfigure}
    \hfill
    \begin{subfigure}[b]{0.32\textwidth}
        \centering
        \includegraphics[width=\textwidth]{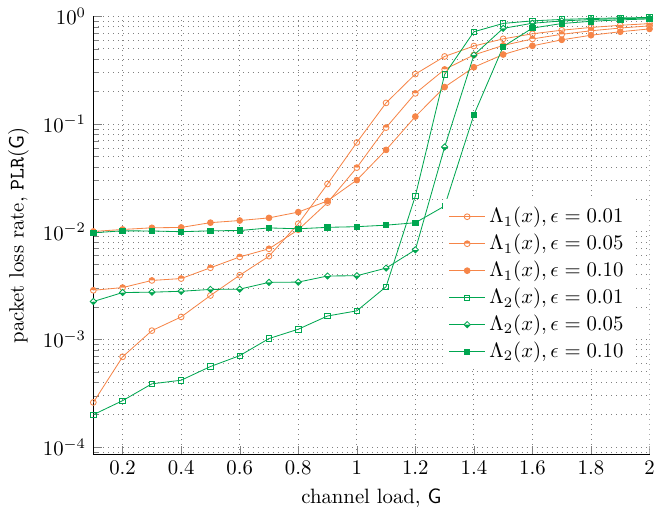}
        \caption{$\slots = 200$ slots, $\kreceivers = 2$ receivers, for degree $\lambda(x)$ distribution and channel erasure probability $\epsilon$}
        \label{fig:PER_dis_epsilon}
    \end{subfigure}
    \caption{Packet loss rate $\plrtot$ as a function of channel load $\load$.}
    \label{fig:mainfig}
\end{figure*}

The performance of the SIC decoding process can be evaluated using tools from code theory on graphs \cite{Liva:IRSA}, \cite{Berioli16}. 
A MAC frame can be represented as a bipartite graph $\mathcal{G} = \{\mathcal{U, S, E} \}$ consisting of a set $\mathcal{U}$ of $\users$ user nodes, a set $\mathcal{S}$ of $\slots$ slot nodes, and a set $\mathcal{E}$ of edges. In the graph representation, an edge $e$ connects the user node $x$ to the slot node $y$ if and only if user $x$ transmitted a copy of its packet over the $y$-th slot.

In the following, we consider the asymptotic scenario, where the frame duration increases indefinitely, i.e.    $\slots \rightarrow \infty$ while maintaining a proportional relationship between the number of users and slots $\users \propto \slots$. This implies that users distribute their replicas across an infinite frame. In this context, we apply the density evolution (DE) algorithm to predict performance of the iterative decoding process.

With DE we can estimate the user decoding failure probability, $\plr$, for a given channel load, $\load$, by iteratively calculating two key probabilities. 
The first, $\q_{i}$, represents the probability that a randomly selected user has not been decoded at the start of the $i$-th SIC iteration. The second, $\p_{i}$, denotes the probability that a randomly selected slot has not been resolved at the beginning of the $i$-th iteration.

Given that $\q_{i}$ reflects a user's unresolved status, 
it can be defined  as the probability that an edge $e$ connecting the user node and the slots node has not been removed in the $i$-th iteration. This occurs when none of the $\ell-1$ replicas was successfully decoded during the previous SIC step, each of with has a probability $\p_{i-1}$. Consequently, we can express this relationship with the following equation
\begin{equation}\label{eq:q_i}
    \q_{i} = \sum_{\ell=1}^{\ell_{\text{max}}} \,   \Lambda_{\ell} \,  \p_{i-1}^{\ell-1}. 
\end{equation}

Instead, $\p_{i}$ reflects a slot's unresolved status, which can be expressed as
\begin{equation}\label{eq:p_i}
    \p_{i} = \sum_{t=1}^{\infty} \,   \tau(t,\epsilon) \,     \p_{i}^{[t]} \end{equation} 
where $\tau(t,\epsilon) $ is the probability that an edge is connected to a slot node of degree $t$  and where $\p_{i}^{[t]}$ is the probability that the edge is not removed at the $i$-th iteration given that is connected to a slot node of degree $t$. 

In other words, $\tau(t,\epsilon) $ represents the probability of having $t$ colliding replicas within the slot,  which 
has been derived in \cite{Liva:IRSA} for a channel without erasures. When considering the effects of the OOF channel, we arrive at the following expression
\begin{equation}\label{eq:pi_t}
    \tau(t , \epsilon) = \frac{ [\bar{\ell} \load (1 - \epsilon)]^{t-1} }{ (t-1)!} e^{-\bar{\ell} \load (1 - \epsilon)}.
\end{equation}
Here, the channel load is scaled by a factor $1 - \epsilon$ to account for the erasures introduced by the OOF channel, reflecting the impact of erased users on the effective channel load.

Let us now focus on $1- \p_{i}^{[t]}$, representing the probability that a slot containing $t$ colliding replicas is solved at the $i$-th iteration. This is determined by two conditions. First, the contribution of all replicas apart from one, i.e. $t-1$, have been canceled in the slot. The probability of this event is $(1- \q_{i})^{t-1}$. Secondly, when all replicas apart from two have been removed in the slot and one has been transmitted at weak power and one at strong power. The probability of this event is $1/2 (t-1) \q_{i} (1- \q_{i})^{t-2}$. Here, the term $1/2$ accounts for the probability of having the two power levels and the term  $(t-1) \q_{i} (1- \q_{i})^{t-2}$ for all possible ways in which all replicas can be canceled from a slot.
By further inserting this result and \eqref{eq:pi_t} into \eqref{eq:p_i} we obtain
\begin{align}
    \p_{i} =  &\sum_{t=1}^{\infty}  \frac{ [\bar{\ell} \, \load (1 - \epsilon)]^{t-1} }{ (t-1)!} e^{-\bar{\ell} \load (1 - \epsilon)} \\ 
   &  \times \Big[ 1  - (1- \q_{i})^{t-1} \big(1  +  \frac{t-1}{2} \frac{ \q_{i} }{1- \q_{i}} \big)   \Big].
\end{align}
With simple manipulations the  expression of  $\p_{(i)}$ becomes
\begin{equation}\label{eq:p_i}
    \p_{i} =  1 - e^{-\bar{\ell} \, \load \, (1 - \epsilon) \q_{i} } \Big[1 + \frac{\bar{\ell} \, \load (1 - \epsilon) \, \q_{i} }{2}  \Big]
\end{equation}
by observing that $\sum_{t=1}^{\infty} \frac{x^t}{t!} = e^{t}$.

The DE algorithm is then performed by iterating equations \eqref{eq:q_i} and \eqref{eq:p_i}.
A pseudocode algorithm for executing DE is provided in Algorithm 1. The inputs include the channel load $\load$, the channel erasure probability $\epsilon$,  replica degree distribution parameters $\Lambda_\ell$ and $\bar{\ell}$, and the maximum number of iterations $ \mathsf{I_{\max}}$ for DE. The initial conditions are set as $\q_{0} = 1$ and $\p_{0} = 1$. The output of the algorithm is the user decoding failure probability at the evaluated channel load, given by  
\begin{equation} \label{eq:pepsilon}
    \plr(\load)= \sum_{\ell=1}^{\ell_{\text{max}}} \,   \Lambda_{\ell} \,  \p_ \mathsf{{I_{\max}}}^{\ell}.
\end{equation} 
In the results, we show that the packet loss rate in \eqref{eq:plr}, defined by density evolution, becomes more accurate as the frame duration increases.

%% file: 5_results.tex
\section{Numerical Results}\label{sec:results}

We now present numerical results for the proposed system, considering the following degree distributions
\[   \Lambda_1(x)  = x^2  \text{ and }  
   \Lambda_2(x)= 0.5465 x^2+ 0.1623 x^3 + 0.2912 x^8.
\]
 The distribution $\Lambda_2(x)$ is noted in literature, for example in \cite{Liva:IRSA}, for its strong performance.
According to the NOMA scheme presented, the system's performance remains independent of the peak power $\power$\footnote{
   Observe that the expressions for  $\plrtot$ and $\plr$ do not depend on the peak power $\power$.}  as long as the parameters $\alpha, p_1$, and $p_2$ are selected as outlined in Sec.~\ref{sec:sys_noma}.  
For simplicity, we assume a unit peak power $\power =1$, and unit duration $\ptime =1$. 
The average energy consumed per node and the energy efficiency values scale simply with $\power$ and $\ptime$.

Let us evaluate the tightness of the derived upper bound of the packet loss rate in   \eqref{eq:plrtot1}
using the output of the DE algorithm with $\mathsf{I}_{\max} = 100$.
For example, consider $\load = 1.2$, $\epsilon=0.05$, and  $\Lambda_1(x)$, then the average number of replicas is  $\bar{\ell} = 2 $. We find that $\p_1 = 0.7811$, $\p_2 = 0.6815, \p_3 = 0.6243.$ and $\p_{100} = 0.4902$ resulting in $\plr(\load = 1.2) = 0.2403$, from \eqref{eq:pepsilon}, which coincides with the simulation results.

In Fig.\ref{fig:DE}, the  $\plrtot(\load)$  of the system  is plotted as a function of the channel load  for the $\Lambda_1(x)$ distribution with $\kreceivers=\{1,2\}$ receivers and channel erasure probability  $\epsilon = 0.05$. The plot illustrates that the trends observed in the simulations (curves with markers) align with the analytical results (dashed curves). It is evident that as the number of slots per frame considered increases, the simulation results converge towards the derived analytical values. As expected, by increasing the number of receivers of one unit the performance of the system improves significantly by decreasing  the $\plrtot$  by one order of magnitude until channel loads of $\load = 1$ [packets/slot].
Note that at low and moderate channel load for one receiver,  the $\plrtot$ is driven primarily  by channel conditions, resulting in an error floor determined by $\epsilon$. We next show how this effect diminishes when considering a larger number of receivers or alternative degree distributions.

Note that the comparison analysis and simulations of $\Lambda_2(x)$ show a tighter match with respect to  $\Lambda_1(x)$ due to the superior performance exhibited by the distribution. However, we do not present these results here due to space constraints.

The upper bound derived enables quick evaluation of trends and the impact of parameter design on performance. It also provides insight into the limits of the best achievable results for a given set of parameters.

Let us now evaluate the impact of an increasing number of receivers.
In this scenario, we consider a MAC frame of $\slots =200$ slots,   OOF erasure  probability of $\epsilon = 0.10$, degree distribution $\Lambda_2(x)$ and different  number of receivers, ranging from $\kreceivers = 1$ to $ \kreceivers = 6$.   
In Fig. \ref{fig:PER_k}, the simulated packet loss rate  $\plrtot$ of the system as a function of the channel load $\load$ is plotted. As expected, performance improves with increasing number of satellite  receivers. Also in this scenario, there is a substantial gain in diversity when moving from one to two satellites. However, the incremental benefit decreases with each additional satellite,  eventually leveling off. Indeed, the impact of $\kreceivers =6$ receivers with respect to $\kreceivers= 5$  becomes minimal. This effect arises because receiver diversity helps mitigate losses introduced by the channel. However, once these channel losses are overcome, further losses are primarily due to failures in SIC decoding. With increased receiver diversity, the impact of the channel is reduced, and packet losses  occur predominantly  due to SIC limitations.


In our next scenario, we show the impact of the degree distributions $\Lambda(x)$ over different values of channel conditions, as shown in Fig.~\ref{fig:PER_dis_epsilon} for  $\mathsf{k}=2$ receivers. Curves with circular markers correspond to the $\Lambda_1(x)$ distribution, while curves without circular markers indicate the $\Lambda_2(x)$ distribution. As expected, the degree distribution significantly influences the packet loss rate performance. For operational channel loads $\Lambda_2(x)$ notably outperforms $\Lambda_1(x)$. This gain increases as the erasures introduced by the channel decrease, e.g. by reducing $\epsilon$. 
However, under severe  channel impairments and lower channel loads (e.g. $\epsilon = 0.10$ and load $\load < 0.6$), the performance difference between the two distributions becomes negligible. This insight may assist  in optimizing the configuration for these conditions and in evaluating other key performance indicators, such as energy efficiency.

In our final scenario, we evaluate the energy efficiency of the system for different network configurations, as shown in Fig.~\ref{fig:energy}. Energy efficiency grows linearly with channel load until reaching a tipping point, corresponding to the maximum throughput that the system can support, after which it declines. For each channel erasure value and number of satellites considered, the distribution $\Lambda_1(x)$ is more energy efficient than $\Lambda_2(x)$. This is mainly due to the fact that the IoT device, on average, transmits fewer replicas with $\Lambda_1(x)$ than with $\Lambda_2(x)$, i.e. $\bar{\ell} = 2$ versus $\bar{\ell} = 3.3271$. Although $\Lambda_2(x)$ achieves higher throughput, the impact on the average replicas transmitted  in \eqref{eq:energy} is more significant. 
Additionally, for a fixed distribution and a single receiver,   the energy efficiency trend shifts as the erasure rate varies.  
This stems from   a shift of the throughput caused by the erased transmissions in the channel. 
As expected, adding a second receiver improves $\eta$, as for a fixed channel load, more users are successfully decoded.  When multiple receivers are considered, energy efficiency further improves under worse channel conditions. This  effect is particularly noticeable at higher channel loads, where decoding failures are more impacted by the SIC algorithm. With degraded channel conditions, each receiver captures a different perspective of the MAC frame, resulting in a higher number of successfully decoded users.

%% file: 6_conclusions.tex
 \vspace{-.2em}
\section{Conclusions}\label{sec:Conclusions}  
We investigated the impact of multi-satellite receivers in an IoT network operating under a NOMA-IRSA scheme within a NTN network. We derived the expressions for the packet loss rate and energy efficiency and formulated the DE analysis, validating it through Monte-Carlo simulations. 
Our results reveal critical trade-offs, highlighting the role of the number of receivers, degree distribution, and channel conditions in evaluating network performance. This analysis provides key insights for designing more resilient and energy-efficient IoT networks in challenging non-terrestrial environments.
\begin{center}
    \begin{figure}[t!]
    \includegraphics[width=0.8\columnwidth]{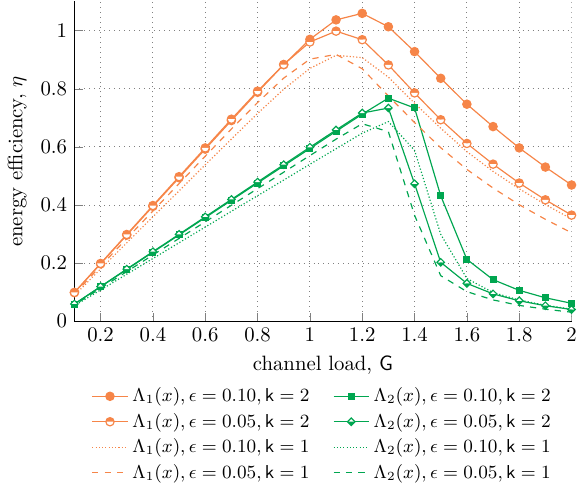}
       \centering 
       \caption{IoT energy efficiency $\eta$ as a function of the channel load $\load$  for $\kreceivers=1,2$ receivers, $\Lambda_1(x)$ and $\Lambda_2(x)$ for different OOF erasure probabilities, and for  $\slots = 200$ slots.}
       \label{fig:energy}
    \end{figure}   
 \end{center}

 \vspace{-1.8em}